\documentclass[aps,prd,preprint,groupedaddress,showpacs,nofootinbib]{revtex4}
\usepackage{graphicx}

\def\kt{\mbox{\boldmath{$k$}}}
\def\zero{\mbox{\boldmath{$0$}}}

\begin{document}

\preprint{DPNU-03-25}
\preprint{hep-ph/0311068}

\title{Analysis of Supersymmetric Effects on $B \to \phi K$ Decays
\\in the PQCD Approach} 

\author{Satoshi Mishima} 
\email[]{mishima@eken.phys.nagoya-u.ac.jp}
\author{A.~I.~Sanda}
\email[]{sanda@eken.phys.nagoya-u.ac.jp}
\affiliation{Department of Physics, Nagoya University, Nagoya 464-8602,
Japan\vspace*{40mm}}


\begin{abstract}
We study the effects of the MSSM contribution on $B\to\phi K$ decays
 using the perturbative QCD approach. In this approach, strong phases
 can be calculated, so that we can predict the values of CP asymmetries
 with the MSSM contribution.  
We predict a large relative strong phase between the
 penguin amplitude and the chromomagnetic penguin amplitude.
If there is a new CP violating phase in the chromomagnetic
 penguin amplitude, then the CP asymmetries may change significantly
 from the SM prediction.
We parametrize the new physics contributions that appear in the Wilson
 coefficients. 
We maximize the new physics parameters up to the point where it is
 limited by experimental constraints.
In the case of the $LR$ insertion, we find that the direct CP
 asymmetries can reach about $85\%$ and the indirect CP asymmetry can
 reach about $-30\%$.
\end{abstract}


\maketitle

%
\section{\label{sec:intro}Introduction}

$B \to \phi K_S$ decay may be useful in the search for new physics
beyond the standard model (SM).  
The time-dependent CP asymmetry of $B^0$ decay into CP eigenstates
$f_{CP}$ can be written as 
$a_{f_{CP}}(t)=A_{f_{CP}}\cos(\Delta M_Bt)+ S_{f_{CP}} \sin(\Delta
M_Bt)$, where
$A_{f_{CP}}$ and $S_{f_{CP}}$ characterize direct
CP violation and indirect CP violation, respectively.
In the SM, both $A_{J/\psi K_S}$ and $A_{\phi K_S}$ vanish,
and both $S_{J/\psi K_S}$ and $S_{\phi K_S}$ must equal to
$\sin(2\phi_1)$.  Any difference between $S_{J/\psi K_S}$ and $S_{\phi
K_S}$ larger than $O(1\%)$ would be a signal for physics beyond
the SM~\cite{Grossman:1996ke}. 
The $B\to \phi K_S$ decay amplitude is induced only at the one-loop
level, so that new physics might contribute to this decay through
quantum effects. At present, BaBar and Belle collaborations have the
following results~\cite{Aubert:2002ic,Abe:2003yu}: 
\begin{eqnarray}
S_{J/\psi K_S} = \left\{
\begin{array}{cl}
 0.741 \pm 0.067\pm 0.034 & (\mathrm{BaBar,\ 81\, fb}^{-1})\;,\\
 0.733\pm 0.057\pm 0.028 & (\mathrm{Belle,\ 140\, fb}^{-1})\;,\\
\end{array}
\right.
\end{eqnarray}
with $A_{J/\psi K_S}=0$. In the $\phi K_S$ mode, they have reported the
following results~\cite{Browder:lp2003,Abe:2003yt}:
\begin{eqnarray}
A_{\phi K_S} &=& \left\{
\begin{array}{cl}
 0.38\pm 0.37\pm 0.12  & \mathrm{(BaBar,\ 110\, fb}^{-1})\;,\\
 -0.15\pm 0.29\pm 0.07 & \mathrm{(Belle,\ 140\, fb}^{-1})\;.\\
\end{array}
\right.
\\
S_{\phi K_S} &=& \left\{
\begin{array}{cl}
 0.45\pm 0.43\pm 0.07 & \mathrm{(BaBar,\ 110\, fb}^{-1})\;,\\
 -0.96\pm 0.50\ {}^{+0.09}_{-0.11} & \mathrm{(Belle,\ 140\, fb}^{-1})\;.\\ 
\end{array}
\right.
\end{eqnarray}
Belle collaboration found a $3.5\sigma$ deviation from the SM
prediction. In contrast with Belle, the BaBar result 
is consistent with the SM. If the Belle result continues to
hold, and both experiments agree, 
then their result is a signal for new physics. 
There have been many papers which studied new physics contributions to
$B\to\phi K_S$ decay amplitude. Some
authors~\cite{Lunghi:2001af,Silvestrini:2002sm,Kane:2002sp}
analyzed the supersymmetric contribution using the mass
insertion approximation, which is a powerful tool for model-independent
analysis of new physics associated with the minimal supersymmetric
standard model (MSSM)~\cite{Hall:1985dx}. 
The new physics contributions come into the Wilson coefficients, which
can be calculated perturbatively~\cite{Gabbiani:1996hi}. The problem is
how to calculate the decay amplitudes with nonperturbative
contributions. To calculate the decay amplitudes, some authors used
naive factorization~\cite{Wirbel:1985ji}, generalized
factorization~\cite{Ali:1997nh}, or QCD
factorization~\cite{Beneke:1999br}.  
Each method is plagued with large theoretical uncertainties.
There are other approaches for this calculation, one of them is
the perturbative QCD (PQCD) approach~\cite{Keum:2000ph}. 
In this paper, we use the PQCD approach and the mass insertion
approximation for an estimation of the MSSM contribution in $B\to \phi
K$ decays. 

\begin{figure}[bt]
\includegraphics{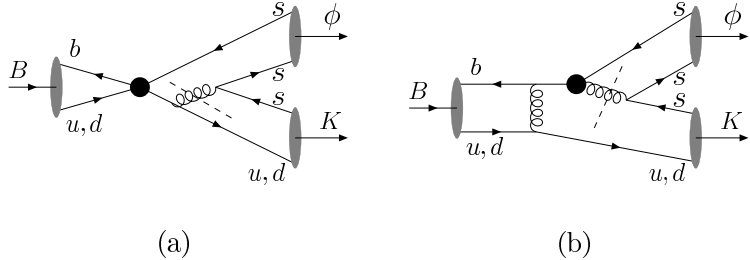}%
\caption{\label{fig:phase}These diagrams generate a large strong
 phase in the PQCD approach. The dashed lines denote the physical
 intermediate states.} 
\end{figure}

The PQCD approach for exclusive $B$ meson
decays is based on the $k_T$ factorization theorem, in which the decay
amplitudes can be separated into perturbative and
non-perturbative parts~\cite{Li:2000hh}. The 
non-perturbative parts are factorized into 
meson wave functions, which are derived from the other methods, for
example, the
light-cone QCD sum rules~\cite{Ball:1998tj,Ball:1998sk,Ball:2003sc}.
A strong phase, which comes from physical intermediate states, is
calculable in the PQCD approach. A large strong phase is induced from an
annihilation diagram such as Fig.~\ref{fig:phase}(a)~\cite{Keum:2000ph}.  
The strong phase mainly comes from the cut on the virtual gluon line.
The source of the strong phase is one of the important differences
between the PQCD approach and other methods.
PQCD has been applied to some hadronic 
two-body $B$ meson decays at leading order in $\alpha_s$, and
the results are consistent with experimental
data except for $\eta^{(\prime)} K$ and
$\pi^0\pi^0$~\cite{Keum:2000ph,Lu:2000em}.  
$B\to \phi K$ decays were also calculated using the PQCD approach in
Refs.~\cite{Mishima:2001pp,Chen:2001pr}. 

Another important difference between PQCD and other methods is
how to calculate magnetic penguin 
diagrams that are induced from the chromomagnetic penguin operator
$O_{8G}$. In many models, the chromomagnetic penguin amplitude is most
sensitive to new physics~\cite{Keum:1999vt,Moroi:2000tk}. However, it is
difficult to calculate the 
chromomagnetic penguin in naive factorization and generalized
factorization, because 
we do not know the magnitude of $q^2$, which is the momentum transferred
by the gluon in the chromomagnetic penguin operator.
Therefore, proponents of these factorization approach
treat $q^2$ as an input parameter, 
so that the result is
directly proportional to the assumed values for $\langle 1/q^2
\rangle$~\cite{Ali:1997nh}. 
In the PQCD and QCD factorization approaches, 
the chromomagnetic penguin amplitudes can be calculated without any
assumption for the value of $q^2$.
We studied the chromomagnetic penguin using the PQCD approach in 
Ref.~\cite{Mishima:2003wm}, and we found that the chromomagnetic penguin
generated a strong phase from the diagram as Fig.~\ref{fig:phase}(b).
In the PQCD approach, $q^2$ is written as
$(1-x_2)x_3M_B^2-|\kt_{2T}-\kt_{3T}|^2$. Here, $x_2$ and $x_3$ are 
momentum fractions of partons in $K$ and $\phi$ mesons, respectively.
$\kt_{2T}$ and $\kt_{3T}$ are transverse momenta of the partons.
In QCD factorization, $q^2$ can be written in terms of the momentum
fraction of partons too, however, they expand the amplitude in power of
$|\kt_{2T}-\kt_{3T}|^2/[(1-x_2)x_3M_B^2]$ and for the leading order they
have $q^2=(1-x_2)x_3M_B^2$.
In this expansion, $q^2$ never vanishes. There is no absorptive part in
the amplitude and the strong phase is not
generated from the chromomagnetic penguin.
The fact that we get large imaginary part implies that this expansion is
not valid. 

The outline of this paper is as follows.
First, we consider the MSSM
contribution in the effective Hamiltonian for $B$ meson decays. 
We present the Wilson coefficients with the MSSM using the mass
insertion approximation.
Next, we briefly review the PQCD approach for the exclusive $B$ meson
decays. We show the result at the leading order in $\alpha_s$ and
how to calculate the chromomagnetic penguin amplitudes.
Furthermore, we calculate the MSSM contribution in $B \to \phi K$
decays with the $LR$, $RL$, $LL$, and $RR$ insertions. 
We consider the both $B^0 \to \phi K_S$ and $B^\pm \to \phi K^\pm$
modes, and calculate the branching ratios, the direct CP
asymmetries, and the indirect CP asymmetry.
Finally, we summarize this study.

%
\section{MSSM Contributions in $B \to \phi K$ Decays}

\subsection{Effective Hamiltonian for $B$ Meson Decays}

We use the effective Hamiltonian in the calculation of $B$ meson
decays~\cite{Buchalla:1995vs}. 
The Hamiltonian is expressed as the convolution of local operators and
the Wilson coefficients.
The effective Hamiltonian with a $b\to s$ transition is given by
\begin{eqnarray}
\hspace{-5mm}
H_{\rm eff}
&=&
\frac{G_F}{\sqrt{2}}
\left[
\sum_{q'=u,c}V_{q's}^* V_{q'b}
\left( C_1(\mu)O_1^{(q')}(\mu)
      + C_2(\mu)O_2^{(q')}(\mu)\right)
\right.
\nonumber\\
& &
\left.
- 
V_{ts}^* V_{tb}
\left(
\sum_{i=3}^{10}C_i(\mu)O_i(\mu)
+
C_{7\gamma}(\mu)O_{7\gamma}(\mu)
+
C_{8G}(\mu)O_{8G}(\mu)
\right) 
\right]
+ {\rm H.c.}
\;, 
\label{hbk}
\end{eqnarray}
where $V_{q's}^*$ and $V_{q'b}$ are the Cabibbo-Kobayashi-Maskawa
matrix elements~\cite{Cabibbo:yz}.
$O_{1-10}$ are local four-fermi operators, 
$O_{7\gamma}$ is the photomagnetic penguin operator, and $O_{8G}$
is the chromomagnetic penguin operator.
The local operators are given by
\begin{eqnarray}
& &O_1^{(q')}
 = (\bar{s}_iq'_j)_{V-A}(\bar{q'}_jb_i)_{V-A}\;,
\;\;\;\;\;\;\;\;
O_2^{(q')}
 = (\bar{s}_iq'_i)_{V-A}(\bar{q'}_jb_j)_{V-A}\;, 
\nonumber \\
& &O_3
 =(\bar{s}_ib_i)_{V-A}\sum_{q}(\bar{q}_jq_j)_{V-A}\;,
\;\;\;\;\;\;\;
O_4
 =(\bar{s}_ib_j)_{V-A}\sum_{q}(\bar{q}_jq_i)_{V-A}\;, 
\nonumber \\
& &O_5
 =(\bar{s}_ib_i)_{V-A}\sum_{q}(\bar{q}_jq_j)_{V+A}\;,
\;\;\;\;\;\;\;
O_6
 =(\bar{s}_ib_j)_{V-A}\sum_{q}(\bar{q}_jq_i)_{V+A}\;, 
\nonumber \\
& &O_7
 =\frac{3}{2}(\bar{s}_ib_i)_{V-A}\sum_{q}e_q(\bar{q}_jq_j)_{V+A}\;,
\;
O_8
 =\frac{3}{2}(\bar{s}_ib_j)_{V-A}\sum_{q}e_q(\bar{q}_jq_i)_{V+A}\;,  
\nonumber \\
& &O_9
 =\frac{3}{2}(\bar{s}_ib_i)_{V-A}\sum_{q}e_q(\bar{q}_jq_j)_{V-A}\;, 
\;
O_{10}
 =\frac{3}{2}(\bar{s}_ib_j)_{V-A}\sum_{q}e_q(\bar{q}_jq_i)_{V-A}\;,  
\nonumber \\
& &O_{7\gamma} 
 =  \frac{e}{8\pi^2} m_b \bar{s}_i \sigma^{\mu\nu}
   (1+\gamma_5) b_i F_{\mu\nu}\;,
\;\;            
O_{8G}    
 =  -\frac{g_s}{8\pi^2} m_b \bar{s}_i \sigma^{\mu\nu}
   (1+\gamma_5)T^a_{ij} b_j G^a_{\mu\nu}\;.
\label{eq:op}
\end{eqnarray} 
Here, $i$ and $j$ are color indices, $q$ is taken to be $u,\;d,\;s$, and
$c$, and $(\bar{q}q)_{V\pm A} = \bar q \gamma_\mu (1\pm \gamma_5)q$. 
We define the covariant derivative as $D_\mu=\partial_\mu
+ig_sT^aA^a_\mu-ieA_\mu$, so that the signs of the magnetic penguin
operators are different between $O_{7\gamma}$ and $O_{8G}$. 
We integrate out the degree of freedom of high energy particles, then
the Wilson coefficients include high energy information.
If we consider the new physics effect on $B$ decays, then we need to
calculate the Wilson coefficients with new physics contributions.

\subsection{MSSM Contribution and Mass Insertion Approximation}

We consider the MSSM contribution for $B\to \phi K$ decays.
Generally, there are new sources of the CP violation and the flavor
changing neutral current (FCNC) in the MSSM, so that there may be direct
CP violations, and it is possible that
$S_{\phi K_S}$ becomes different from $S_{J/\psi K_S}$.
Since we do not want our computation to depend on specific SUSY models,
we use the mass insertion approximation to calculate the Wilson
coefficients with the MSSM. In the mass insertion
approximation, the FCNC effect appears in the squark propagators
through the off-diagonal elements in the squark mass matrices. 
The decay amplitudes are expanded in terms of $(\delta^d)_{ij}=
(V_d^\dagger m^2_{\tilde d}V_d)_{ij}/m^2_{\tilde q}$, where
$m_{\tilde d}^2$ is 
the squared down-type squark mass matrix,
$m_{\tilde q}$ is an average squark mass, and $V_d$ is the matrix which
diagonalizes the down-type quark mass matrix. Of course, we must
consider the region of $|(\delta^d)_{ij}| < 1$.
For example, a transition of a right-handed fermion to a left-handed
fermion is parameterized by $(\delta^d_{LR})_{ij}$.
There are four mass insertions: $(\delta^d_{LL})_{ij}$,
$(\delta^d_{RR})_{ij}$, $(\delta^d_{LR})_{ij}$, and
$(\delta^d_{RL})_{ij}$. The $b\to s$ transition is induced from the
gluino-squark loop, the chargino-squark loop, and the neutralino-squark
loop. In this study, we consider only the gluino contribution, which is
dominant in many models. 
The Wilson coefficients for penguin and magnetic penguin are given
by 
\begin{eqnarray}
C_3^{\rm NP} (M_S)
 &\simeq&
 -\frac{\sqrt{2} \alpha_s^2}{4G_F V_{ts}^* V_{tb} m_{\tilde{q}}^2}  
 (\delta_{LL}^d)_{23} \left[ -\frac{1}{9} B_1(x) - \frac{5}{9} B_2(x)
 - \frac{1}{18} P_1(x) -\frac{1}{2} P_2(x) \right]
\;,\nonumber\\
C_4^{\rm NP}(M_S)
 &\simeq&
 -\frac{\sqrt{2} \alpha_s^2}{4G_F V_{ts}^* V_{tb} m_{\tilde{q}}^2} 
(\delta_{LL}^d)_{23}
\left[ -\frac{7}{3} B_1(x) + \frac{1}{3} B_2(x) + \frac{1}{6} P_1(x)
+\frac{3}{2} P_2(x) \right]
\;,\nonumber\\
C_5^{\rm NP} (M_S)
 &\simeq&
 -\frac{\sqrt{2} \alpha_s^2}{4G_F V_{ts}^*V_{tb} m_{\tilde{q}}^2} 
 (\delta_{LL}^d)_{23}
\left[ \frac{10}{9} B_1(x) + \frac{1}{18} B_2(x) - \frac{1}{18} P_1(x)
-\frac{1}{2} P_2(x) \right]
\;,\nonumber\\
C_6^{\rm NP} (M_S)
 &\simeq&
 -\frac{\sqrt{2} \alpha_s^2}{4G_F V_{ts}^* V_{tb} m_{\tilde{q}}^2} 
 (\delta_{LL}^d)_{23}
\left[ -\frac{2}{3} B_1(x) + \frac{7}{6} B_2(x) + \frac{1}{6} P_1(x)
+\frac{3}{2} P_2(x) \right]
\;,\nonumber\\
C_{7\gamma}^{\rm NP} (M_S)
&\simeq&
\frac{\sqrt{2} \alpha_s \pi}
{6G_F V_{ts}^*V_{tb}  m_{\tilde{q}}^2}
\left[
(\delta_{LL}^d)_{23}\frac{8}{3} M_3(x)+
(\delta_{LR}^d)_{23}\frac{m_{\tilde{g}}}{m_b}
\frac{8}{3} M_1(x) \right]
\;,\nonumber\\
C_{8G}^{\rm NP} (M_S)
&\simeq&
\frac{\sqrt{2} \alpha_s \pi}
{2G_F V_{ts}^* V_{tb} m_{\tilde{q}}^2}
\left[
(\delta_{LL}^d)_{23}\left( \frac{1}{3} M_3(x) + \!3
			       M_4(x)\right)
\right.\nonumber\\
& & \left. \ \ \ \ \ \ \ \ \ \ \ \ \ \ \ \ \ \ \ 
+ 
(\delta_{LR}^d)_{23}\frac{m_{\tilde{g}}}{m_b}
\left(\frac{1}{3} M_1(x) + 3 M_2(x)\right)\right]
\label{eq:cmp}
\;,
\end{eqnarray}
at the first order in the mass insertion
approximation~\cite{Gabbiani:1996hi}\footnote{The signs of
$(\delta_{LR}^d)_{23}$ terms are opposite from those in
Ref.~\cite{Kane:2002sp} and the same as those in
Ref.~\cite{Gabbiani:1996hi}.}. 
Here, $M_S$ is the SUSY scale, and $x =
m^2_{\tilde{g}}/m^2_{\tilde{q}}$, where $m_{\tilde{g}}$ is the gluino
mass. $B(x)$, $P(x)$ and $M(x)$ are the loop functions, which are
calculated from box diagrams and penguin
diagrams~\cite{Gabbiani:1996hi,Harnik:2002vs}.  
New physics contributions induce additional operators, which are
obtained from Eq.~(\ref{eq:op}) by exchanging $L$ and $R$.
The penguin coefficients $C_3^{\rm NP}(M_S)-C_6^{\rm NP}(M_S)$ depend
only on $\delta^d_{LL}$, and the 
magnetic-penguin coefficients $C_{7\gamma}^{\rm NP}(M_S)$ and $C_{8G}^{\rm
NP}(M_S)$ depend on both $\delta^d_{LL}$ and $\delta^d_{LR}$. 
It is noted that the $\delta^d_{LR}$ terms have a chiral enhancement
factor $m_{\tilde{g}}/m_b$. This factor is important when we
constraint the parameters from the branching ratio for $B\to X_s
\gamma$, as we will return it on later.

%
\section{$B\to \phi K$ Decays in PQCD Approach}

\begin{figure}[bt]
\includegraphics{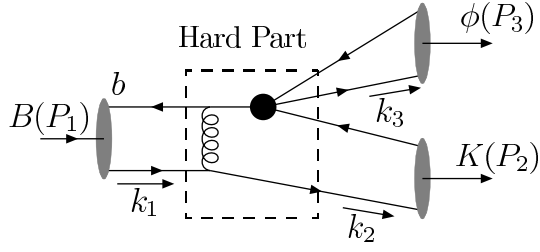}%
\caption{\label{fig:phik_momenta}A leading-order diagram in $B\to \phi
 K$.}  
\end{figure}

Let us briefly review the PQCD approach for exclusive $B$ meson decays. 
The PQCD formalism is based on the factorization of decay amplitudes
into a product of long-distant physics, which is identified with meson
wave functions, and short-distant physics~\cite{Lepage:1979zb}.
The meson distribution amplitudes are universal in the processes under
consideration, and they are determined from experiments and/or other
theoretical methods: the light-cone QCD sum rules, lattice calculations,
etc. Process dependence is exhibited in the short-distant part. 
For example, let us consider $B\to \phi
K$. Figure.~\ref{fig:phik_momenta} shows the PQCD quark-level diagram
for this decay. At the black blob, the decay $\bar b\to \bar s s \bar s$
takes place. The $s\bar s$ pair forms a $\phi$ meson. The other $s$
quark recoils against the $\phi$ meson carrying almost $m_b/2$
momentum. In order for the spectator quark to form a $K$ meson together
with the fast moving $s$ quark, it has to exchange a gluon in order to
change its momentum from $k_1$ to $k_2$.

We consider the $B\to\phi K$ decays.
In the light-cone coordinates, the $B$ meson momentum $P_1$, the $K$
meson momentum $P_2$, and the $\phi$ meson momentum $P_3$ are taken to
be 
\begin{eqnarray}
P_1=\frac{M_B}{{\sqrt 2}}(1,1,\zero_T)\;,\;\;
P_2=\frac{M_B}{{\sqrt 2}}(1-r_\phi^2,0,\zero_T)\;,\;\;
P_3=\frac{M_B}{{\sqrt 2}}(r_\phi^2,1,\zero_T)\;,
\end{eqnarray}
where $r_\phi = M_\phi/M_B$.
Here, we consider the $B$ meson to be at rest, and
the $K$ meson mass is 
ignored.  
The momenta of partons $k_1$, $k_2$, and $k_3$ defined in
Fig.~\ref{fig:phik_momenta} are written as 
\begin{eqnarray}
k_1=(0,x_1P_1^-,\kt_{1T})\;,\;\;
k_2=(x_2P_2^+,0,\kt_{2T})\;,\;\;
k_3=(0,x_3P_3^-,\kt_{3T})\;.
\end{eqnarray}
A hard part in Fig.~\ref{fig:phik_momenta} has two propagators:
\begin{eqnarray}
\frac{1}{(k_1-k_2)^2}\ \cdot\
\frac{1}{(P_1-k_2)^2-M_B^2}
&\simeq& 
\frac{1}{-x_1x_2M_B^2-|\kt_{1T}-\kt_{2T}|^2}\ \cdot\
\frac{1}{-x_2M_B^2-|\kt_{2T}|^2}\;.
\end{eqnarray}
If we neglect transverse momenta of partons, then a singularity arises
from the end-point region of parton momenta since the hard part is
proportional to $1/x_1x_2^2$.  
Therefore, we conclude that the transverse components are present and
they cannot be ignored. 
Retaining $k_T$ for partons, large double logarithms appear through
radiative corrections. 
The resummation of those double logarithms leads
to the Sudakov factor~\cite{Botts:kf}.
The Sudakov factor suppresses the end-point of the parton momentum and
the large transverse separation between a quark and an antiquark in
mesons. Therefore, it guarantees a perturbative calculation of the hard 
part~\cite{Li:1992nu}.  
The other double logarithms appear from the end-point
region of the parton momenta. Their resummation, which is the so-called
threshold resummation, leads to another factor in the hard
part~\cite{Li:2001ay}. 
This factor also ensures the absence of the end-point singularities in
PQCD, and the arbitrary cutoffs used in QCD factorization 
are not necessary. 
A typical decay amplitude for $B\to\phi K$ can be expressed as the
convolutions of a hard part, meson wave functions, and the Wilson
coefficient, in the space
of $x_i$ and $b_i$, where $b_i$ is the conjugate variable to
$k_{iT}$~\cite{Chang:1996dw}. 
\begin{eqnarray}
{\cal M} &=& \int_0^1 dx_1 dx_2 dx_3 
\int_0^\infty db_{1} db_{2} db_{3}\, 
\Phi_K(x_2,b_{2})\, e^{-S_K(x_2,b_2,t)}\, 
\Phi_\phi(x_3,b_{3})\, e^{-S_\phi(x_3,b_3,t)}
\nonumber\\
& &\hspace{17mm}\times
C\left(t\right)  H(x_1,x_2,x_3,b_{1},b_{2},b_{3},t)\, J(x_1,x_2,x_3)\,
\Phi_B(x_1,b_{1})\, e^{-S_B(x_1,b_1,t)}\;.
\label{eq:ktfac}
\end{eqnarray}
Here, $\Phi_B$, $\Phi_K$, and $\Phi_\phi$ are the meson wave functions,
and $H$ is the hard part. $S_B$, $S_K$, and $S_\phi$ denote the Sudakov
factors, and $J$ denotes the threshold factor.  
The scale $t$, which characterizes the hard part, is of order of
$\sqrt{\bar\Lambda M_B}$ where $\bar\Lambda=M_B-m_b$. 
The formula in Eq.~(\ref{eq:ktfac}) is the typical amplitude for an
exclusive $B$ meson decays based on the $k_T$ factorization.

\subsection{$B\to \phi K$ in the Standard Model} 

\begin{figure}[bt]
\includegraphics{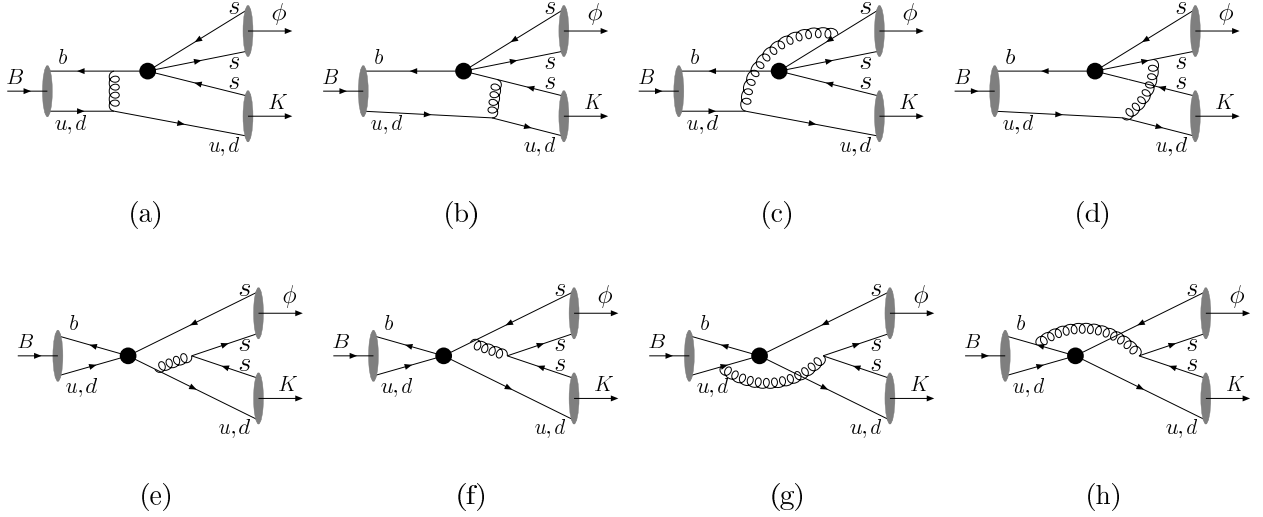}%
\caption{\label{fig:pqcd_phik}Leading order diagrams for $B\to\phi K$
 decays. The ellipses denote the meson wave functions, and the thick
 points denote the local operators.} 
\end{figure}

For $B\to \phi K$ decays, we calculate the diagrams shown in
Fig.~\ref{fig:pqcd_phik} at leading order in the PQCD approach.
The diagrams (a) and (b) are dominant contributions, and the diagrams
(e) and (f) generate a large strong
phase~\cite{Keum:2000ph}. 
Except for a small tree diagram contribution in the $B^\pm \to \phi
K^\pm$ decay amplitude, both $B^0\to \phi K^0$ and $B^\pm \to \phi
K^\pm$ decay amplitudes get contributions from pure penguin graphs.
For $B$ meson, we use the model wave function
\begin{eqnarray}
\phi_B(x,b) &=& N_B
x^2 (1-x)^2\exp\left[-\frac{1}{2}\left(\frac{xM_B}{\omega_{B}}\right)^2
-\frac{\omega_{B}^2 b^2}{2}\right] \;,
\end{eqnarray}
where $b$ is the conjugate space of $k_T$.
$\omega_B$ is the shape parameter to be $0.40\pm 0.04$ and $N_B$ is
the normalization constant~\cite{Kurimoto:2001zj}. 
For $K$ and $\phi$ mesons, we use the wave functions
that were calculated using the light-cone QCD sum
rules~\cite{Ball:1998tj,Ball:1998sk,Ball:2003sc}. 
The formulas of the decay amplitudes for $B\to\phi K$ are shown in
Ref.~\cite{Mishima:2001pp}\footnote{For $K$ meson, the
moments $a_1$ and $a_2$ were recalculated in Ref.~\cite{Ball:2003sc}.}.
We get the following numerical results within the SM:
\begin{eqnarray}
\mathrm{Br}(B^0\to \phi K^0)
&=&
 (8.5{}^{+3.0}_{-2.0}\pm 2.6) \times 10^{-6}\;,
\\
\mathrm{Br}(B^\pm\to \phi K^\pm)
&=&
(9.3{}^{+3.1}_{-2.1}\pm 2.8) \times 10^{-6}\;.
\end{eqnarray}
The values for various parameters that we use in this
calculation are presented in Appendix~\ref{sec:app_par}. 
Here, the first error is estimated from the shape parameter $\omega_B$
in the $B$ meson wave function, and the second error comes from 
higher-order contributions. We expect that the higher-order
contributions are about $30\%$.
The theoretical errors are reduced in CP asymmetries, because errors
associated with wave functions cancel out between the denominator and
the numerator. 
The current experimental data are given
by~\cite{Aubert:2003hz,unknown:2003jf}:
\begin{eqnarray}
\mathrm{Br}(B^0\to \phi K^0) &=& \left\{
\begin{array}{cl}
(8.4 {}^{+1.5}_{-1.3}\pm 0.5)\times 10^{-6} & (\mathrm{BaBar})\;,\\
(9.0 {}^{+2.2}_{-1.8} \pm 0.7) \times 10^{-6} & (\mathrm{Belle})\;,\\
\end{array}
\right.
\\
\mathrm{Br}(B^\pm\to \phi K^\pm) &=& \left\{
\begin{array}{cl}
(10.0 {}^{+0.9}_{-0.8} \pm 0.5)\times 10^{-6} & (\mathrm{BaBar})\;,\\
(9.4 \pm 1.1 \pm 0.7) \times 10^{-6} & (\mathrm{Belle})\;.\\
\end{array}
\right.
\end{eqnarray}
The predicted branching ratios in PQCD are consistent with the
experimental data.

\subsection{Chromomagnetic Penguin and New Physics}

\begin{figure}[bt]
\includegraphics{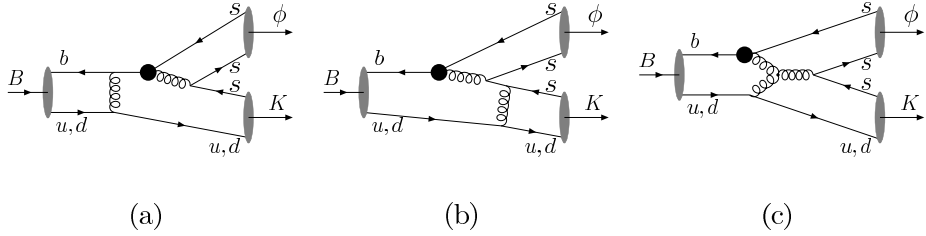}%
\caption{\label{fig:pqcd_mag}chromomagnetic penguin diagrams in
 $B\to\phi K$ decays. We show only dominant diagrams in the
 chromomagnetic penguin amplitude.}  
\end{figure}

The chromomagnetic penguin operator $O_{8G}$ in Eq.~(\ref{eq:cmp})
plays an important role in the estimation of new physics
contribution. The chromomagnetic penguin diagrams are shown in
Fig.~\ref{fig:pqcd_mag}. 
We show only dominant diagrams in the chromomagnetic
penguin amplitude~\cite{Mishima:2003wm}.
Since there must be at least one hard gluon 
emitted by the spectator quark, the chromomagnetic penguin amplitudes
are of next-to-leading order in $\alpha_s$ in the PQCD
formalism. Obviously, there are many other higher-order diagrams that
must be considered simultaneously. 
Here we limit ourselves to computing only the nonvanishing
leading-order terms and leave the higher-order terms for future
computation. 
As we pointed out before new physics contribution comes in through
the chromomagnetic penguin in spite of the fact that the leading term
is ${\cal O}(\alpha_s^2 )$.  
Of course, regular penguin amplitudes may also contain new physics and
they are ${\cal O}(\alpha_s)$, and we include them. 
In summary, we calculate the amplitude
\begin{eqnarray}
{\cal A} = 
C^{\rm SM}_{1-10}
\otimes H\left(\alpha_s\right)
\ +\ 
C^{\rm NP}_{3-6}
\otimes H\left(\alpha_s\right)
\ +\ C^{\rm NP}_{8G}
\otimes H\left(\alpha_s^2\right)\;.
\end{eqnarray}
As we will see below
that these penguin amplitudes and chromomagnetic penguin amplitudes
give comparable contributions.

%
\section{Numerical Analysis}

In this section, we estimate the MSSM effect on the branching ratios
and the CP asymmetries for both $B^0\to\phi K^0$ and $B^\pm\to\phi
K^\pm$ decays. We take single mass insertion: one of the $LR$, $RL$,
$LL$, and $RR$ insertions, which are parametrized by $\delta_{LR}$,
$\delta_{RL}$, $\delta_{LL}$, and $\delta_{RR}$, respectively. First, we
constrain them from the branching ratio for $B\to X_s\gamma$, which is
an inclusive decay mode with the $b\to s$
transition~\cite{Kagan:1998ym}. This mode is theoretically very clean.
The theoretical prediction with in the context of the SM 
agrees with experimental data within errors, so that this mode will give
meaningful constraint on any new physics we might introduce.
Next, we apply the constrained parameters to $B\to\phi K$
decays. We calculate the branching ratios and the direct and indirect CP
asymmetries with the MSSM contribution.

\subsection{Constraint from $B\to X_s \gamma$}

\begin{figure}[tb]
\includegraphics{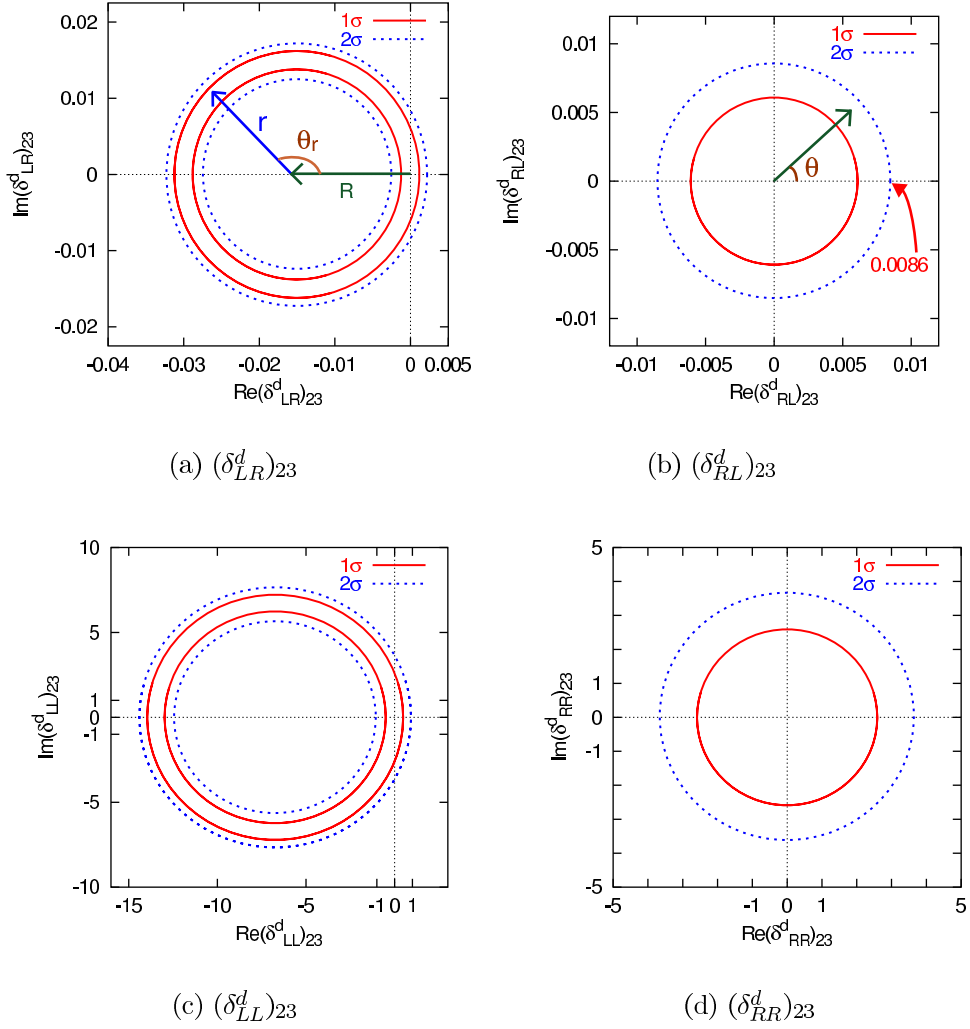}%
\caption{\label{fig:bsgamma}The allowed region for
 $(\delta^d_{LR})_{23}$, $(\delta^d_{RL})_{23}$, 
 $(\delta^d_{LL})_{23}$, and $(\delta^d_{RR})_{23}$ constrained from
 $\mathrm{Br}(B\to X_s\gamma)$. The solid lines denote the region in
 the $1\sigma$ error, and the dotted lines are the $2\sigma$ error. We
 take $m_{\tilde g}$ and $m_{\tilde q}$ to be $500$ GeV.
 In the analysis of $B\to\phi K$, we take the following parametrization:
 $(\delta^d_{LR})_{23} = R + r e^{i\theta_r}$ and $(\delta^d)_{23} =
 |(\delta^d)_{23}|e^{i\theta}$ for the $RL$, $LL$, and $RR$ insertions.}  
\end{figure}

We constrain the mass insertion parameters from ${\rm Br}(B\to X_s
\gamma)$.  The experimental result is $\mathrm{Br}(B\to X_s \gamma) =
(3.3\pm 0.4)\times 10^{-4}$~\cite{Hagiwara:fs}, and we take it with the
$2\sigma$ error to constrain the parameters. The results for each
insertion are shown in Fig.~\ref{fig:bsgamma}.
In this calculation, we take the gluino mass
$m_{\tilde g}$ and the squark mass $m_{\tilde q}$ to be $500$ GeV.
The constraints for $(\delta^d_{LR})_{23}$ and  $(\delta^d_{RL})_{23}$
are strong, while those for $(\delta^d_{LL})_{23}$ and 
$(\delta^d_{RR})_{23}$ are very weak. 
This is because amplitudes with $(\delta^d_{LR})_{23}$ and
$(\delta^d_{RL})_{23}$ are enhanced by a $m_{\tilde g}/m_b$ factor.

\subsection{$B\to \phi K$ Decays with the MSSM}

We calculate the MSSM effect on $B^0\to\phi K^0$ and $B^\pm\to\phi
K^\pm$ decays with $LR$, $RL$, $LL$, and $RR$ insertions. 
In the case of the $LR$ insertion, we parametrize
$(\delta^d_{LR})_{23}$ as
\begin{eqnarray}
(\delta^d_{LR})_{23}
= R + r e^{i\theta_r}\;,
\end{eqnarray}
where $R$ is a constant, and $r$ and $\theta_r$ are parameters as shown
 in Fig.~\ref{fig:bsgamma}(a). 
In the SM,
$(\delta^d_{LR})_{23}$ is equal to $0$, that is, $r=-R$ and
$\theta_r=0$. 
Since the constraints for 
$(\delta^d_{LL})_{23}$ and $(\delta^d_{RR})_{23}$
from $\mathrm{Br}(B\to X_s\gamma)$ are very weak,
we take the arbitrary bounds $|(\delta^d_{LL})_{23}|<1$ and
$|(\delta^d_{RR})_{23}|<1$ 
in order to use the mass insertion approximation. 
Both the $LL$ and $RR$ insertions have the same contribution to
 $B\to\phi K$ decays.
In the case of the $RL$ and $LL$($RR$) insertions, we use the angle
$\theta$ in Fig.~\ref{fig:bsgamma}(b) as a parameter:
\begin{eqnarray}
(\delta^d)_{23} = |(\delta^d)_{23}|e^{i\theta}\;.
\end{eqnarray}
In the following analysis, we scan all values on the allowed
region in Fig.~\ref{fig:bsgamma}(a) for the $LR$ insertion.
For the $RL$ and $LL$($RR$) insertions, we take some specific values
$|(\delta^d_{RL})_{23}| = 0.001$ or $0.0086$, and
$|(\delta^d_{LL(RR)})_{23}| = 0.5$ or $1.0$.  

The $B_s-\overline{B_s}$ mixing may also be affected by the MSSM
contribution~\cite{Silvestrini:2002sm}, so that we 
have examined it in order to constraint the mass insertion
parameters. The current experimental data is $\Delta M_s > 14.4\
\mathrm{ps}^{-1}$ (at 95\% C.L.)~\cite{Stocchi:2002yi}.
The values of $\Delta M_s$ is not very sensitive to the presence of the
$LR$ and $RL$ insertions. 
In the case of the $LL$ and $RR$ insertions, their allowed regions are
reduced somewhat but these insertions in the $B\to\phi K$
amplitudes is small so that it does not affect our analysis.  

\begin{figure}[bt]
\includegraphics{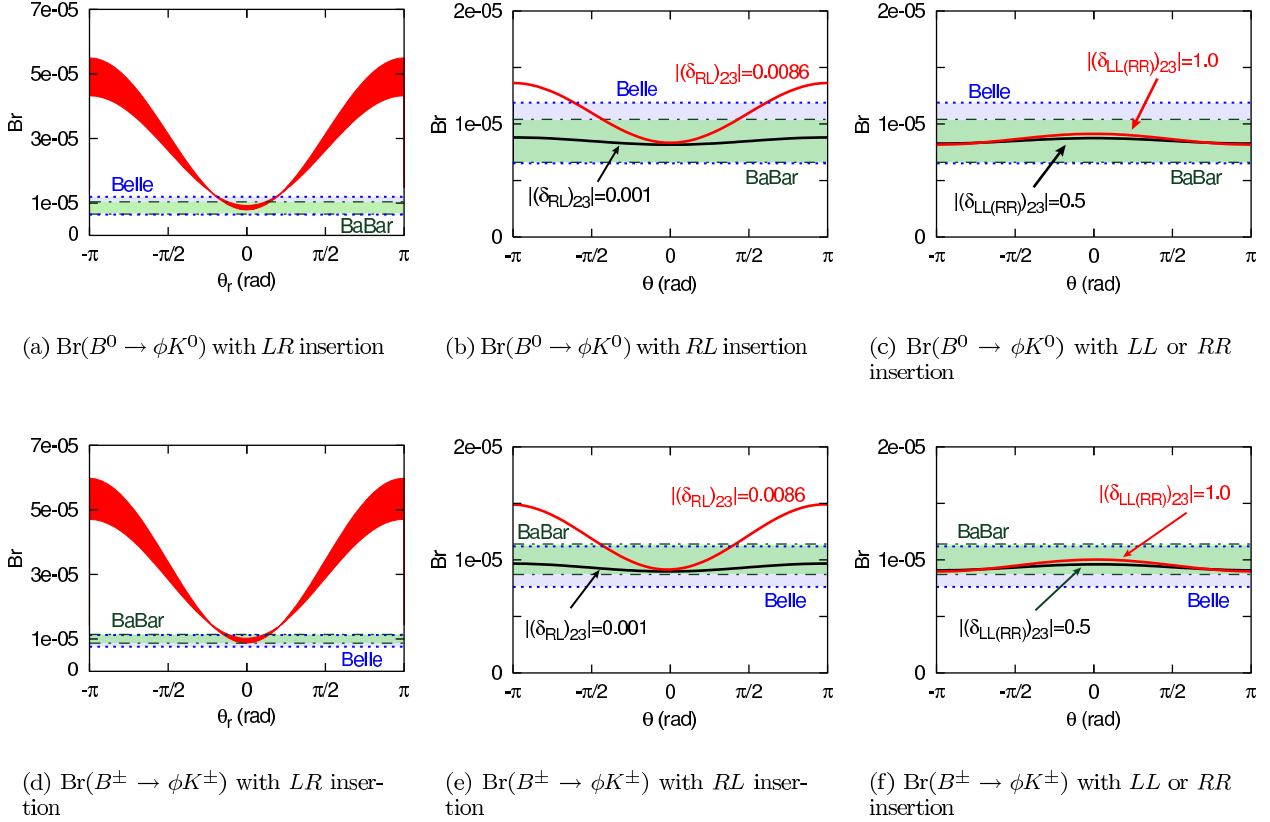}%
\caption{\label{fig:phik_br}The branching ratios with the MSSM
 contribution. We take $m_{\tilde g}$ and $m_{\tilde q}$ to be $500$
 GeV. The dot-dashed lines and the dotted lines are the experimental data
 with the $1\sigma$ error by BaBar and Belle, respectively.}
\end{figure}

Possible MSSM modification for $B\to\phi K$ branching ratios are shown
in Fig.~\ref{fig:phik_br}.  
As it can be seen from the figures, the $LR$ insertion may give large
effect on the branching ratios for $B\to \phi K$ decays, while the
contributions from $RL$
and $LL$($RR$) insertions are small. In the PQCD approach, there are
large theoretical uncertainties in the calculation of the branching
ratios. Therefore, it is difficult to obtain meaningful constraints from
the branching ratios for $B\to \phi K$ decays. In the case of the
$LR$ insertion, we suppose $|\theta_r|$ is less than $\pi/2$. In the
case of the other insertions, we cannot constrain the parameters from
the branching ratios.

\begin{figure}[bt]
\includegraphics{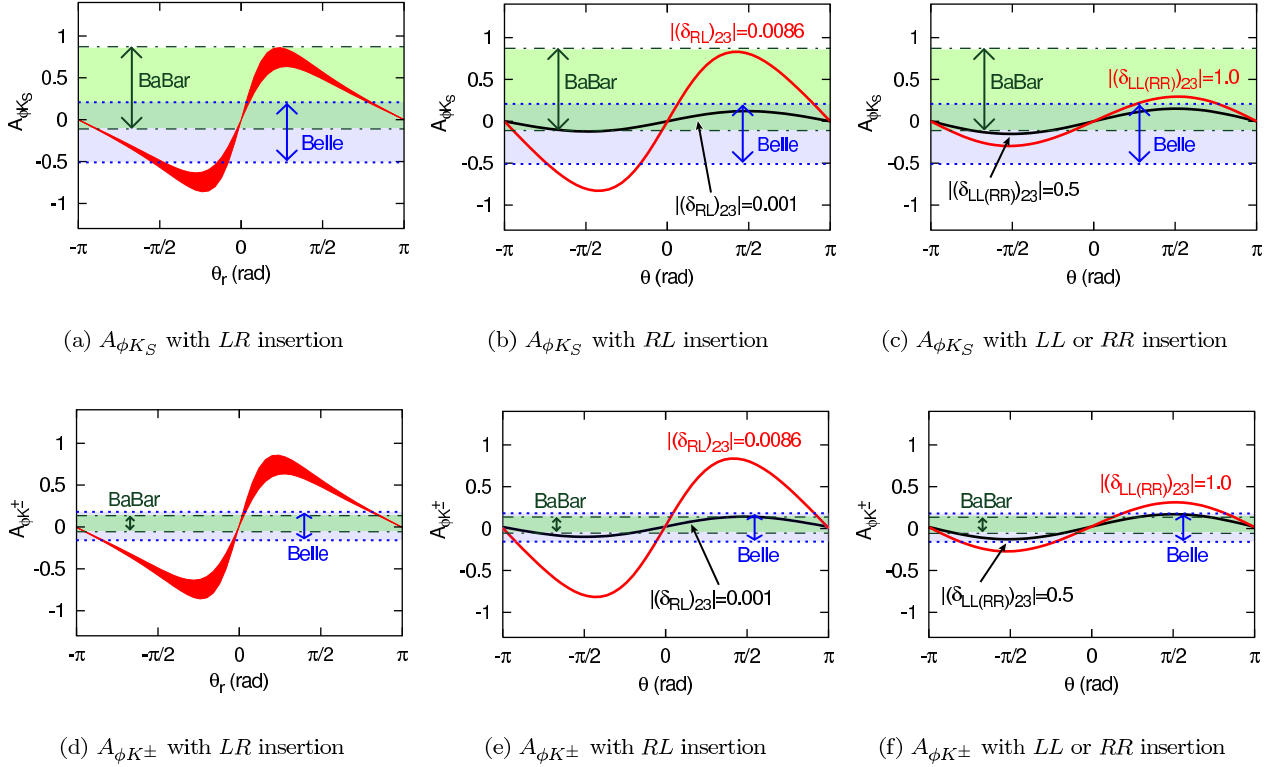}%
\caption{\label{fig:phik_a}Direct CP asymmetries with the MSSM
 contribution. We take $m_{\tilde g}$ and $m_{\tilde q}$ to be $500$
 GeV. The dot-dashed lines and the dotted lines are the experimental data
 with the $1\sigma$ error by BaBar and Belle, respectively.} 
\end{figure}

The results of the direct CP asymmetries are shown in
Fig.~\ref{fig:phik_a}.  
The experimental data of the direct CP asymmetry for the charged mode are
${\cal A}_{\phi K^\pm}=0.04\pm 0.09\pm 0.01$
(BaBar~\cite{Aubert:2003hz}) and $0.01\pm 0.12\pm 0.05$
(Belle~\cite{unknown:2003jf}). In the SM,  
the direct CP asymmetries are almost 0, since $B\to\phi K $ decays are
penguin dominant processes. In the case of the $LR$ insertion, the
direct CP asymmetries can reach about $85\%$. These asymmetries are
generated from the interference between penguin amplitudes in the
SM and chromomagnetic penguin amplitudes in the MSSM. Since
the relative strong phase between those amplitudes is large, it might be 
possible to get the large direct CP asymmetries in the PQCD approach. 
It is noted that the direct CP asymmetry in the neutral mode is almost
the same as one in the charged mode. This result from the fact that the
chromomagnetic penguin contributions as well as the SM contributions
are almost the same in both modes.  
Since the current experimental data of the direct CP asymmetry in the
charged mode is small, we expect that one in the neutral mode will
be also small. 

\begin{figure}[bt]
\includegraphics{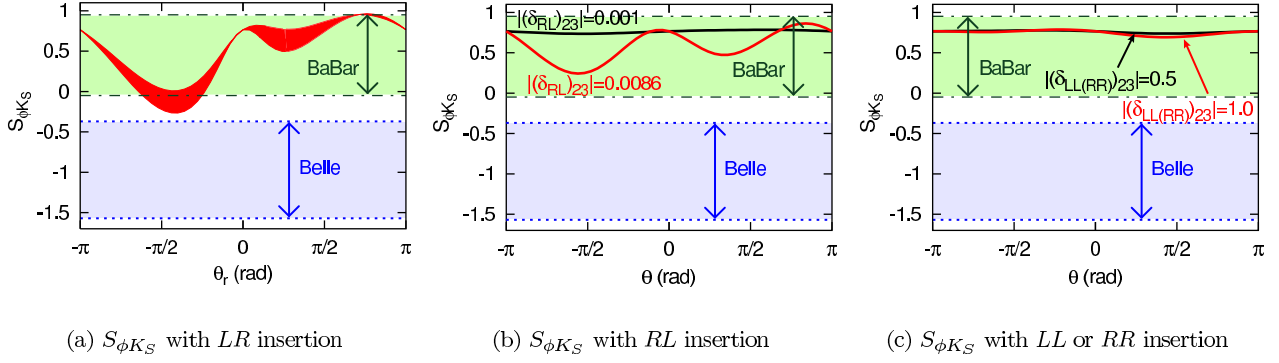}%
\caption{\label{fig:phik_s}The result of $S_{\phi K_S}$ with the MSSM
 contribution. We take
 $m_{\tilde g}$ and $m_{\tilde q}$ to be $500$ GeV.
 The dot-dashed lines and the dotted lines are the experimental data 
 with the $1\sigma$ error by BaBar and Belle, respectively.}
\end{figure}

\begin{figure}[bt]
\includegraphics{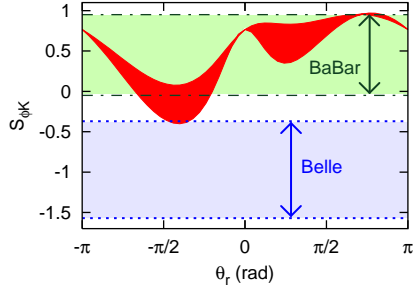}%
\caption{\label{fig:phik_s_2000} As an example of the extreme case, we
 take $m_{\tilde g}$ to be $200$ GeV and $m_{\tilde q}$ to be 
 $2$ TeV in the case of the $LR$ insertion. The dot-dashed lines and the
 dotted lines are the experimental data with the $1\sigma$ error by BaBar
 and Belle, respectively.}
\end{figure}

Next, we consider the indirect CP asymmetry in $B^0\to\phi K_S$ decay.
The results are shown in Fig.~\ref{fig:phik_s}. 
The current BaBar data is consistent with the SM prediction,
however, Belle data is not. Our result is
$S_{\phi K_S} \geq -0.28$ in the $LR$ insertion case. 
If we change the gluino mass and the squark mass, while the ratio is
fixed, then $S_{\phi K_S}$ does not change. If we fix the 
squark mass and take a heavier gluino mass, then $S_{\phi K_S}$
becomes larger. 
If we fix the gluino mass and take a heavier squark mass, then
$S_{\phi K_S}$ becomes somewhat smaller, that is, new physics
contribution becomes larger, and $S_{\phi K_S}$ remains
almost constant over $m_{\tilde q}$ to be a few TeV. Here, we
constrained $(\delta^d_{LR})_{23}$ from ${\rm Br}(B\to 
X_s\gamma)$, so that if we take a larger $m_{\tilde q}$, then
$(\delta^d_{LR})_{23}$ increases and $C_{8G}^{NP}$ increases slightly.
In order to study the effect of the mass on $S_{\phi K_S}$, we take 
an extreme case where the 
gluino mass is $200$ GeV and the squark mass is $2$ TeV. 
In this case, $(\delta^d_{LR})_{23}$ is still ${\cal O}(10^{-2})$, and
the mass insertion approximation can be used.
The result is shown in Fig.~\ref{fig:phik_s_2000}.
Even in this case, the $S_{\phi K_S}$ can only reach about $-40\%$. 
Therefore, it is difficult to explain the current Belle data by the new
physics we considered here.

Finally, we comment one of the differences between our results and that
of QCD factorization.
A negative $S_{\phi K_S}$ in Fig.~\ref{fig:phik_s}(a) implies a negative
$A_{\phi K_S}$ in Fig.~\ref{fig:phik_a}(a). In an analysis using QCD
factorization, a negative $S_{\phi K_S}$ implies that 
$A_{\phi K_S}(=-C_{\phi K_S})$ is positive in contrast to our
results~\cite{Kane:2002sp}. 
It is caused by the difference in the origin of the strong phase in PQCD
and QCD factorization.

\section{Conclusion}

$B\to \phi K$ is one of the most important decay modes in the search for
new physics. In this paper, we estimated MSSM contribution in the
$B\to\phi K$ decays using the PQCD approach. In PQCD, strong phases are
calculable, and we can predict CP asymmetries. We considered the single
mass insertions $(\delta_{LR}^d)_{23}$, $(\delta_{RL}^d)_{23}$,
$(\delta_{LL}^d)_{23}$, and $(\delta_{RR}^d)_{23}$, and constrain 
them from ${\rm Br}(B\to X_s\gamma)$. We found that the $LR$ insertion
may change the branching ratios and the CP asymmetries in $B\to \phi K$
significantly. The effect of the $RL$ insertion is somewhat smaller than
the $LR$ insertion, and that of the $LL$ and $RR$ insertions is little. 

In the case of the $LR$ insertion, $A_{\phi K}$ can reach
about $\pm 0.85$ in both neutral and charged modes, and $S_{\phi K_S}
\geq -0.28$. The direct CP asymmetries arise from the interference
between  penguin amplitudes in the SM and chromomagnetic penguin
amplitudes in the MSSM. In PQCD, there is a large relative strong phase
between them, so that the direct CP asymmetries may be large depending
on the new physics parameters. 
As in Figs.~\ref{fig:phik_s}(a) or \ref{fig:phik_s_2000} and
Fig.~\ref{fig:phik_a}(a) indicate, our result is
incompatible with the current Belle data.
However, the current Belle result is not in agreement with the current
BaBar result, so that we need more data to arrive at a definite
conclusion.  
Finally, it must be noted that the direct CP asymmetry in
the neutral mode has the same tendency as one in the charged mode,
because the chromomagnetic penguin contributions as well as the SM
contributions are almost the same in both modes.

%
\begin{acknowledgments}
We would like to thank Y.Y.~Keum, E.~Kou, T.~Kurimoto, H-n.~Li, 
 M.~Matsumori, and Y.~Shimizu for useful comments and discussions. 
S.~M. acknowledges support from the Research Fellowships of the Japan
 Society for the Promotion of Science for Young Scientists
 (No.13-01722).   
A.~I.~S. acknowledges support from the Japan Society for the Promotion
of Science, Japan-US collaboration program, and a grant from Ministry of
Education, Culture, Sports, Science and Technology of Japan.
\end{acknowledgments}
%
%
\appendix

\section{\label{sec:app_par}Input Parameters}
The parameters that we used in this study are as
follows~\cite{Hagiwara:fs}:
$|V_{ts}| = 0.0412$,\
$|V_{tb}| = 1.0$,\
$|V_{us}| = 0.2196$,\
$|V_{ub}| = 0.0036$,\
$\phi_3 = 80^\circ$,\
$M_B = 5.28\; {\rm GeV}$,\ 
$M_K =0.49 \;{\rm GeV}$,\ 
$M_{\phi}=1.02\; {\rm GeV}$,\
$m_b=4.8 \; {\rm GeV}$,\ 
$m_t = 174.3 \; {\rm GeV}$,\
$f_{B}= 190\; {\rm MeV}$,\ 
$f_{K} = 160\; {\rm MeV}$,\ 
$f_{\phi} = 237 \;{\rm MeV}$,\
$f_{\phi}^T = 220 \;{\rm MeV}$,\
$\tau_{B^0}=1.54\times 10^{-12}\;{\rm sec}$,\ 
$\tau_{B^\pm}=1.67\times 10^{-12} \;{\rm sec}$,\
$\Lambda_{\rm QCD}^{(4)}=0.250\;{\rm GeV}$, and 
$m_{0K} = M_K^2/(m_d+m_s) = 1.7$ GeV.
\\

%
%

%
\end{document}